\documentclass[showpacs,amsmath,amssymb]{revtex4}

\usepackage{amsfonts}
\usepackage{amsmath}
\usepackage{amssymb}
\usepackage{graphicx}
\usepackage{dcolumn}
\usepackage{times}
\usepackage{color}
\usepackage{epstopdf}

\begin{document}

\author{X.N. Maintas}
\affiliation{Department of Physics, University of Athens, GR-15771 Athens, Greece}
\author{C.E. Tsagkarakis}
\affiliation{Department of Physics, University of Athens, GR-15771 Athens, Greece}
\author{F.K. Diakonos}
\affiliation{Department of Physics, University of Athens, GR-15771 Athens, Greece}
\author{D.J. Frantzeskakis}
\affiliation{Department of Physics, University of Athens, GR-15771 Athens, Greece}

\title{Nonlinear Schr\"{o}dinger solitons in massive Yang-Mills theory
and partial localization of Dirac matter}





\begin{abstract}
We investigate the classical dynamics of the massive $SU(2)$ Yang-Mills field in the framework of multiple scale perturbation theory. We show analytically that there exists a subset of solutions having the form of a kink soliton, modulated by a plane wave, in a linear subspace transverse to the direction of free propagation. Subsequently, we explore how these solutions affect the dynamics of a Dirac field possessing an $SU(2)$ charge. We find that this class of Yang-Mills configurations, when regarded as an external field, leads to the localization of the fermion along a line in the transverse space. Our analysis reveals a mechanism for trapping $SU(2)$ charged fermions in the presence of an external Yang-Mills field indicating the non-abelian analogue of Landau localization in electrodynamics.
\end{abstract}

\maketitle

\section{Introduction}

Over the last decades, the classical dynamics of Yang-Mills (YM) field theory has been thoroughly investigated in the literature, both in Minkowski and in Euclidean space (see, e.g., Ref.~\cite{Smilga01} and references therein). The motivation for this study has been mainly the effort to understand the vacuum structure of non-abelian gauge theories like Quantum Chromodynamics (QCD). In a spatially homogeneous description, one can show that the YM classical dynamics possesses a chaotic component attributed to the nonlinear form of the YM self-interaction \cite{Smilga01,Savvidi81}. Generalizing to the case of inhomogeneous solutions, the conformal structure of the YM Lagrangian and the associated absence of a characteristic scale does not permit the presence of localized solutions \cite{Coleman77}, and complicated patterns with fractal characteristics may appear \cite{Wellner92}. Recently, it has been argued that classical Yang-Mills solutions may have impact on the properties of the quantum gauge fields. In particular, in Ref.~\cite{Frasca06}, it was shown that periodic solutions of a special choice for the YM field configuration (Smilga's choice \cite{Smilga01}) after quantization lead to a description of the gauge field propagator compatible with the calculations performed in lattice gauge theories.

On the other hand, localized inhomogeneous solutions could permit a particle interpretation of the YM-field, which may be relevant for several applications where quasi-particles are involved. Such a scenario appears, for example, when the YM-field is coupled to a condensate, breaking spontaneously the underlying gauge symmetry, or when the YM-field itself condensates at particular thermodynamic conditions. In these cases the gauge field can acquire a mass introducing a {\it scale} in the YM-theory and bypassing the restrictions of the Coleman theorem \cite{Coleman77}. This  allows for spatially inhomogeneous localized classical solutions -- at least at the level of an effective theory.

In the present work, we follow this line of thoughts trying to explore the space of classical solutions in massive $SU(2)$ Yang-Mills theory. Our primary interest is to display the capacity of the theory in terms of possible classical dynamical behavior, as well as the influence of the choice for the YM-field initial configuration on this dynamics. In particular we will show that at a given combination of scales the classical Yang-Mills theory contains the non-linear Schr\"{o}dinger equation regime. We start our considerations with a Langrangian describing the interaction of the Yang-Mills field with a scalar field. Then we assume, at the level of the Langrangian, that the scalar field is constant and we remain with a massive Yang-Mills theory. The effect of the spatio-temporal fluctuations of the scalar field is considered in \cite{Maintas2011}. As a next step, making a choice similar to Smilga's \cite{Smilga01}, we are able to construct within the framework of a multi-scale perturbation theory a class of solutions which are localized along a line in the plane transverse to the momentum of the gauge field.

Furthermore, we study the dynamics of Dirac fields in the presence of such a gauge field configuration, considering the latter as an external classical field. We show that the Dirac field becomes bound in the subspace where the external gauge field is localized.

The paper is organized as follows: in section II we present the Lagrangian of the considered $SU(2)$ YM field theory, we discuss the multiple scale approach used to solve the corresponding equations of motion and we obtain the associated solutions for the gauge field. We also give an interpretation of the involved parameters. In section III we use the solution found in section II as an external field for the Dirac dynamics of an
$SU(2)$-charged matter field.  Finally we end up, in section IV, with a summary and perspectives of our work.

\section{Soliton-like solutions in the massive Yang-Mills dynamics}

We start our analysis by considering the Lagrangian describing the interaction of the $SU(2)$ Yang-Mills field $A^{a}_{\mu}$ with a charged scalar field $\Phi=\left( \begin{array}{c} \Phi_1 \\ \Phi_2 \end{array} \right)$:
\begin{equation}
\mathcal{L}= - {1\over 4}F^a_{\mu\nu}F^{a \mu\nu}+ \left[ \left( \partial_{\mu} + i g A^{a}_{\mu} \tau_a \right) \Phi \right]^{\dagger}
\left[ \left( \partial^{\mu} + i g A^{a,\mu} \tau_a \right) \Phi \right]-V[\Phi^{\dagger} \Phi]
\label{eq:eq0}
\end{equation}
where $g$ is a dimensionless coupling and $V[\Phi^{\dagger} \Phi]$ is the self-interaction potential of the scalar field which we need not to specify more. We only assume that the potential possesses at least one stable equilibrium point. As usual, we use greek letters to denote the space-time components and latin letters to denote the Lie group components of the YM fields. For the $SU(2)$ case $a,b,c$ take the values $1,2,3$. Let us now further assume that the scalar field is constant (independent of space-time) and equal to a value corresponding to a stable equilibrium point of $V$. Then the Lagrangian (\ref{eq:eq0}), up to the constant term $V[\Phi^{\dagger} \Phi]$ which can be neglected, becomes:
\begin{equation}
\mathcal{L}= - {1\over 4}F^a_{\mu\nu}F^{a \mu\nu}+{1\over 2} M^2_{ab} A^{\mu a}A_\mu^b
~~~;~~~F^a_{\mu\nu} = \partial_{\mu}A^{a}_{\nu}-\partial_{\nu}A^{a}_{\mu}
- g \varepsilon_{abc}A^{b}_{\mu}A^{c}_{\nu},
\label{eq:eq1}
\end{equation}

\noindent
In Eq.~(\ref{eq:eq1}) $M_{ab}$ is the mass matrix of the YM field components which is diagonal in the group indices $M_{11}=M_{22}=M_{33}=m_g=\vert \Phi_1 \vert^2 + \vert \Phi_2 \vert^2$. The corresponding evolution equations are given by:
$$(\Box\delta _{ab}  + M^2_{ab})A_\nu^b - \partial_\nu \partial^\mu A_{a \mu} + g \varepsilon_{abc}\left[A_\mu^b\partial_\nu A^{\mu c}-A_\nu^c\partial^\mu A_\mu^b-2A_\mu^b\partial^\mu A_\nu^c\right] $$

\begin{equation}
-g^2\left[ A_{a \nu} A_\mu^b A^{\mu b} - A_\nu^b A_{a \mu} A^{\mu b}\right] =0,
\label{eq:eq2}
\end{equation}

\noindent
where $\delta _{ab}$ and $\varepsilon_{abc}$ are the Kronecker delta and the full antisymmetric tensor in $SU(2)$ space, respectively. We use the multiple-scale perturbation theory (see, e.g., Ref.~\cite{jefkaw}) to solve the nonlinear Eqs.~(\ref{eq:eq2}): first, we introduce the new space-time independent variables, $X^{\mu_n}$, as well as the partial derivatives thereof:
$$X^{\mu_n} = \epsilon ^n x^\mu \quad ,~~~~\epsilon <<1$$

\begin{equation}
\partial_\mu\rightarrow \partial_{\mu_0} + \epsilon \partial_{\mu_1} +\epsilon^2 \partial_{\mu_2} + \epsilon^3 \partial_{\mu_3} + \ldots, \label{eq:eq3}
\end{equation}

\noindent
and we assume that the corresponding field variables are expanded into an asymptotic series of the form:
\begin{equation}
A_\mu^a \rightarrow \epsilon  A_\mu^a(1) + \epsilon^2  A_\mu^a(2) + \epsilon^3  A_\mu^a(3) +\ldots,
\label{eq:eq4}
\end{equation}

\noindent
where $\epsilon$ is a formal small parameter (connected to the kink soliton amplitude and inverse width -- see below). Substituting the above expressions into the equations of motion, and equating coefficients of the same powers of $\epsilon$, we obtain a set of equations from which $A_\mu^a(k)$ ($k=0,1,2,\cdots$) can be successively determined. Notice that each field $A_\mu^a(k)$ is to be determined so as to be bounded (nonsecular) at each stage of the perturbation.

In order to solve the evolution equations arising at various orders in $\epsilon$, one can make an appropriate choice for the gauge field components, allowing for their decoupling -- at least in the lowest orders in the perturbation expansion. Here, we will use the following configuration for the gauge fields:
\begin{eqnarray}
A_1^1,A_2^2=\mathcal{O}(\epsilon^1), \nonumber \\
A_3^1,A_3^2,A_0^1,A_0^2=\mathcal{O}(\epsilon^2), \nonumber \\
A_3^3,A_0^3,A_1^2,A_2^1,A_1^3,A_2^3=\mathcal{O}(\epsilon^4),
\label{eq:eq5}
\end{eqnarray}
\noindent
which allows us to decouple the corresponding equations of motion up to the order
$\mathcal{O}(\epsilon^3)$. This configuration is in fact a generalization of the Smilga's choice \cite{Smilga01} for spatial non-homogeneous
fields (see Appendix A).

The resulting simplified equations for the component $A_k^k$ ($k=1,2$) are given as follows:
\begin{eqnarray}
\mathcal{O}(\epsilon): (\Box_0 + m_g^2 + \partial_{k_0}^2)A_k^k(1)=0, \phantom{aaaaaaaaaaaaaaaaaaaaaaaaaaaaa}
\label{eq:eq6}\\
\mathcal{O}(\epsilon^2): (\Box_0 +m_g^2 +\partial_{k_0}^2) A_k^k(2) +  2\partial_{\mu_0} \partial^{\mu_1}A_k^k(1)=0, \phantom{aaaaaaaaaaaaaaa}
\label{eq:eq7}\\
\mathcal{O}(\epsilon^3):(\Box_0 + m_g^2 + \partial_{k_0}^2 )A_k^k(3) + (2\partial_{\mu_0} \partial^{\mu_1} + 2 \partial_{k_0} \partial_{k_1})A_k^k(2)-\partial_\xi^2 A_k^k(1)    \nonumber \\
 +(\Box_1 +\partial_{k_1}^2+ 2\partial_{\mu_0} \partial^{\mu_2} + 2 \partial_{k_0} \partial_{k_2}  )A_k^k(1) +g^2 S_k=0,
\label{eq:eq8}
\end{eqnarray}
\noindent
where we have used the notation:
$$\xi=\epsilon (x+y)~~,~~S_1\equiv A_1^1(1)A_2^2(1)A_2^2(1)~~,~~S_2 \equiv A_2^2(1)A_1^1(1)A_1^1(1).$$
Here we should note that there is no summation over repeated latin indices in Eqs.~(\ref{eq:eq6}-\ref{eq:eq8}). The equations of the remaining components are obtained in a similar way. Equation~(\ref{eq:eq8}) still contains a coupling between $A_1^1$ and $A_2^2$, due to the nonlinear term, which can be resolved using the further assumption: $A_1^1 \equiv A_2^2$ \cite{Smilga01}.

Equations~(\ref{eq:eq6}-\ref{eq:eq8}) can be solved self-consistently, leading to the following equations satisfied by
the unknown $A_1^1(1)$ component:
\begin{eqnarray}
O(\epsilon):&  A_1^1(1)=f_1^1(X_{\mu_1},X_{\mu_2},\ldots;1) e^{-i \tau} ~+~c.c. \quad,~~~ \tau \equiv k_0 t-k_z z,
\label{eq:eq9} \\
O(\epsilon^2):& f(1)\equiv f_1^1(X_{\mu_1},X_{\mu_2},\ldots;1)=f(\epsilon (x+y),\ldots;1),
\label{eq:eq10} \\
O(\epsilon^3):&(\Box_1 + 2\partial_{\mu_0} \partial^{\mu_2} + \partial_{X_1}^2)A_1^1(1)-\partial_\xi^2A_1^1(1) +g^2 A_1^1(1)A_1^1(1)A_1^1(1)=0,\phantom{aaa}
\label{eq:eq11}
\end{eqnarray}
where $k_0^2-k_z^2=m_g^2$. After some simple algebraic manipulations, the nonlinear evolution equation~(\ref{eq:eq11}) takes the usual form of a nonlinear Schr\"{o}dinger (NLS) equation with a repulsive (self-defocusing) nonlinearity (due to $g^2 >0$ in the nonlinear term):
\begin{equation}
 -2\partial_{\xi}^2 f(1) - 2i k_0 {\partial f(1)\over \partial T_2} + 3g^2 f(1)|f(1)|^2 =0, \label{eq:eq12}
\end{equation}
\noindent
which has been studied extensively in various branches of physics and, especially, in nonlinear optics
\cite{kivpr} and atomic Bose-Einstein condensates \cite{djf}. The above NLS equation possesses a stationary kink-type (alias ``dark'') soliton solution \cite{zsd}, given by:

\begin{equation}
A\equiv\epsilon A_1^1(1)=\epsilon A_2^2(1)=\epsilon{2 \alpha \over \sqrt{3}g}\tanh(\alpha \xi) e^{-i(\tau+F_0^2 T_2)}~+~c.c. \label{eq:eq13}
\end{equation}
\noindent
where $T_2=\epsilon^2 t $ and $ \alpha=\sqrt{{k_o \over 2}}F_0$.
Details on the derivation of Eq.~(\ref{eq:eq12}) are provided in Appendix A.

In Fig.~1 we show a plot of the solution (\ref{eq:eq13}) using the parameter values: $\alpha=53$~MeV $\epsilon=0.1$,
$k_0=550$~MeV, and $F_0=3.2$~MeV$^{1/2}$. It can be seen that the obtained form is characterized by a free propagation in $z$-direction and a kink-soliton profile in the $\xi$-direction, with $\xi=\epsilon (x+y)$.

\begin{figure}[ht]
\includegraphics[scale=0.65]{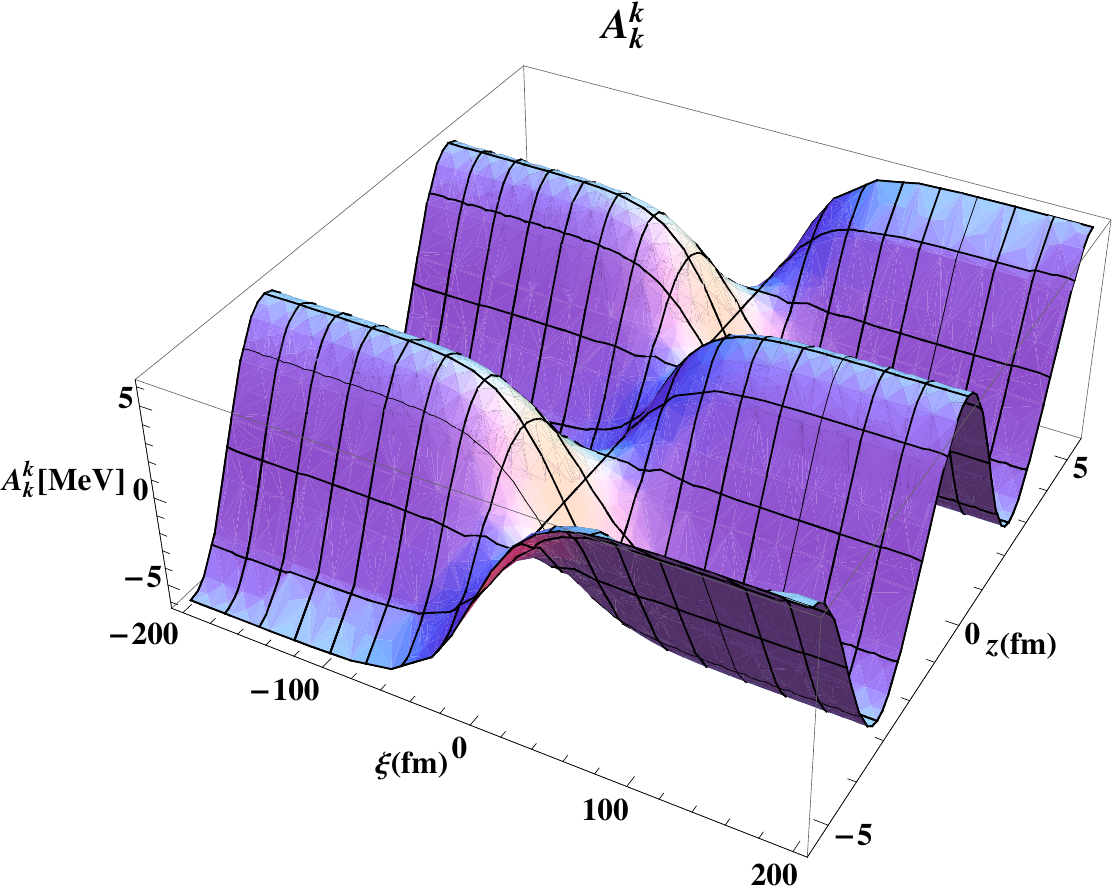}
\caption{The kink-type solution of Eq.~(\ref{eq:eq13}) for the components of the $SU(2)$ gauge field, using the parameter values: $\alpha=53$~MeV, $\epsilon=0.1$,
$k_0=550$~MeV, and $F_0=3.2$~MeV$^{1/2}$.
\label{fig1}}
\end{figure}

It is obvious that Eq.~(\ref{eq:eq12}), due to the presence of a first derivative in time, breaks the Lorentz invariance of the initial Lagrangian density; this is in accordance to the assumptions made to obtain the consistent solution (\ref{eq:eq13}) decomposing space-time in two inequivalent subspaces
($(x,y)$ and $(z,t)$). This property is inevitably expected to hold for gauge field solutions varying over a finite space interval. Additionally, gauge invariance is violated from the very beginning due to the presence of the gauge field mass term. However, the validity of the solution (\ref{eq:eq13}) is restricted to specific space-time scales and, therefore, there is no apparent contradiction with first principles.

After suitable rescaling in order to introduce dimensionless quantities, we have checked the validity of the solution (\ref{eq:eq13}) through numerical integration of eqs.~(\ref{eq:eq2}). Adapting the choice (\ref{eq:eq5}) for the configuration of the gauge fields we concentrate on the equations of motion for the diagonal components $A^k_k$ ($k=1,2$). The results of our numerical treatment in $1+1$ dimensions for $A^1_1(\Xi,t)$ ($\Xi=x+y$) \footnote{Notice that $A^2_2(\Xi,t)=A^1_1(\Xi,t)$ holds for all considered times in accordance with our choice (see \cite{Smilga01})} is shown in the contour plot of Fig.~2. The solution (\ref{eq:eq13}) holds for more than 100 field oscillations indicating its remarkable stability and supporting the validity of our perturbative scheme.

\begin{figure}[ht]
\includegraphics[scale=0.65]{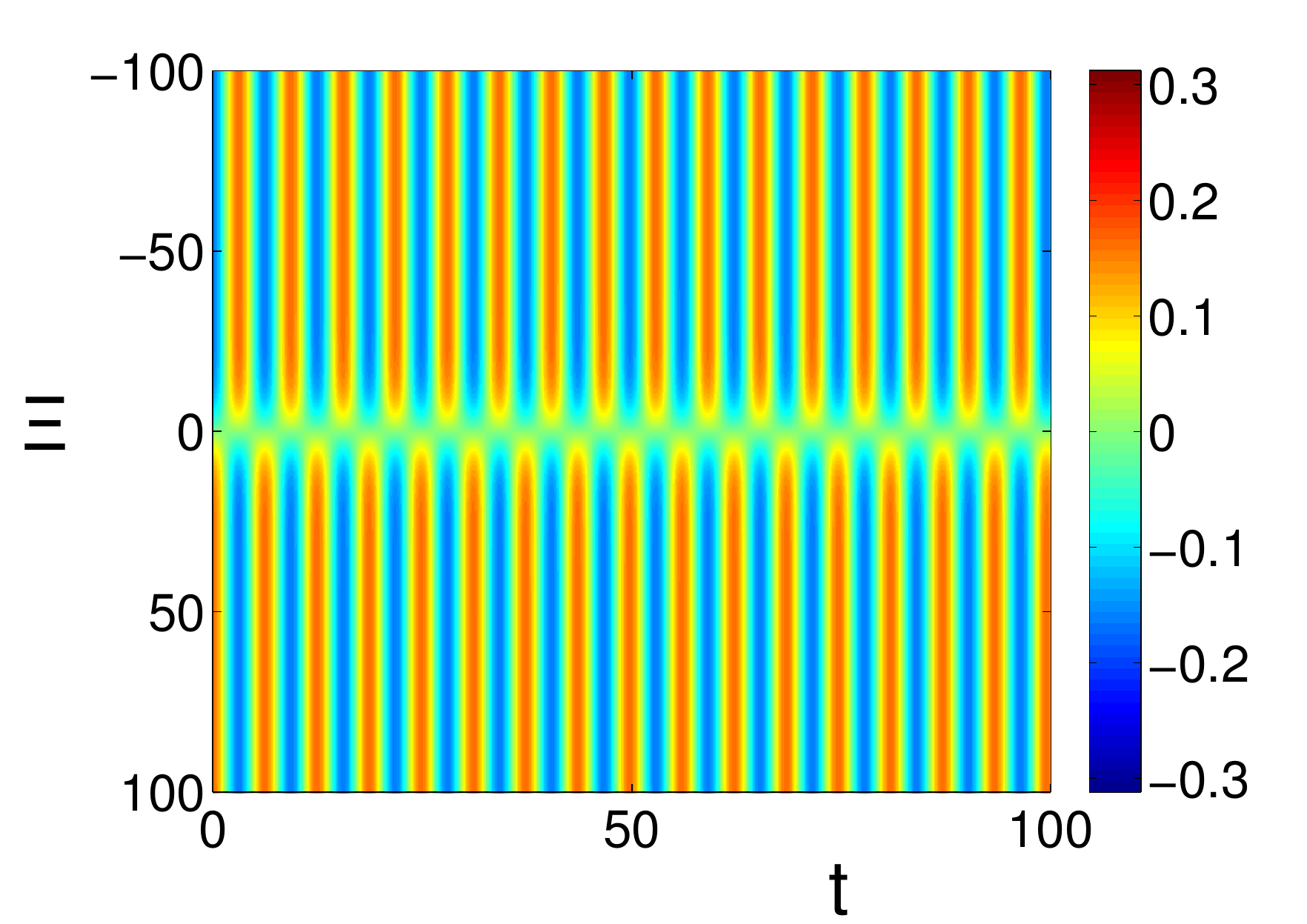}
\caption{Contour plot of the numerical solution for $A^1_1(\Xi,t)$ using as initial condition the analytically obtained form given by eq.~(\ref{eq:eq13}). The length scale is $m_g^{-1}$. We have also used $\epsilon=0.1$.}
\label{fig2}
\end{figure}

\section{Partial localization of Dirac matter}

In this section we will investigate the dynamics of an $SU(2)$ charged Dirac field in the presence of an external gauge field which has the form found in Eq.~(\ref{eq:eq13}). The corresponding Dirac equation is written as follows:
\begin{equation}
(i \gamma^{\mu} D_{\mu}-m)\Psi=0~~~;~~~D_{\mu}=\partial_{\mu}+i \frac{g}{2}A^{a}_{\mu} \sigma^a,
\label{eq:eq14}
\end{equation}
where $\sigma_a$ ($a=1,2,3$) are the Pauli spin matrices, $\gamma_{\mu}$ ($\mu=0,1,2,3$) are the Dirac matrices, and $\Psi=\left(\begin{array}{c}\Psi_1 \\ \Psi_2 \end{array}\right)$ is the $SU(2)$ doublet for the fermionic field. For the fermionic mass matrix $M$ we assume a diagonal form with $m_{11}=m_{22}=m_f$. Due to the non-abelian character of the gauge group, the equations describing the dynamics of the two charged fields $\Psi_1$ and $\Psi_2$, after expanding (\ref{eq:eq14}) and substituting Eq.~(\ref{eq:eq13}) for the non-abelian gauge field, take the following coupled form:
\begin{equation}
(i \gamma^{\mu} \partial_{\mu}-m_f) \Psi_1={1 \over 2} g  A (\gamma^1- i \gamma^2)\Psi_2
\label{eq:eq15}
\end{equation}

\begin{equation}
(i \gamma^{\mu} \partial_{\mu}-m_f) \Psi_2={1 \over 2} g  A (\gamma^1+ i \gamma^2)\Psi_1
\label{eq:eq16}
\end{equation}

\noindent
Taking into account that the expression (\ref{eq:eq13}) for the gauge field is non-covariant, it is consistent to consider the dynamics implied by Eqs.~(\ref{eq:eq15}-\ref{eq:eq16}) in the non-relativistic limit. For that purpose, it is necessary to write the bispinors $\Psi_1$ and $\Psi_2$ in terms of their components.
In that regard, we introduce the following notation:
\begin{equation}
\Psi_1 = \left( \begin{array}{c}\chi_1 \\ \phi_1 \end{array} \right) = \left( \begin{array}{c}\chi_{11} \\ \chi_{12} \\ \phi_{11} \\ \phi_{12} \end{array} \right)~~~;~~~\Psi_2 = \left( \begin{array}{c} \chi_2 \\ \phi_2 \end{array} \right) = \left( \begin{array}{c}\chi_{21} \\ \chi_{22} \\ \phi_{21} \\ \phi_{22} \end{array} \right)
\label{eq:eq17}
\end{equation}
Applying the standard procedure (see, e.g., Ref.~\cite{Bjorken78}) for obtaining the non-relativistic limit of Eqs.~(\ref{eq:eq15}-\ref{eq:eq16}) (details of the calculations are given in Appendix B), we find the following set of coupled Schr\"{o}dinger-type equations for the fermionic components $\widetilde{\chi}_{ij}$ (where ${\chi}_{ij}=\exp(-i m_f t)\widetilde{\chi}_{ij}$):

\begin{equation}
i \partial_t \widetilde{\chi}_{11}+ {\nabla^2 \over 2m_f}\widetilde{\chi}_{11}=
-{1+i \over 2m_f}g\epsilon \partial_\xi(A\widetilde{\chi}_{21}),
\label{eq:eq18}
\end{equation}

\begin{equation}
i \partial_t \widetilde{\chi}_{12}+{\nabla^2 \over 2m_f}\widetilde{\chi}_{12}=
-{i k_z \over 2m_f}g(\partial_\tau A)\widetilde{\chi}_{21}-\epsilon{1+i \over 2m_f}gA \partial_\xi\widetilde{\chi}_{22}+{g^2 \over 2m_f}A^2\widetilde{\chi}_{12},
\label{eq:eq19}
\end{equation}

\begin{equation}
i \partial_t \widetilde{\chi}_{22}+ {\nabla^2 \over 2m_f}\widetilde{\chi}_{22}=
\epsilon{1-i \over 2m_f}g\partial_\xi(A\widetilde{\chi}_{12}),
\label{eq:eq20}
\end{equation}

\begin{equation}
i \partial_t \widetilde{\chi}_{21}+{\nabla^2 \over 2m_f}\widetilde{\chi}_{21}
={i k_z \over 2m_f}g(\partial_\tau A)\widetilde{\chi}_{12}+\epsilon{1-i \over 2m_f}gA \partial_\xi\widetilde{\chi}_{11}+{g^2 \over 2m_f}A^2\widetilde{\chi}_{21},
\label{eq:eq21}
\end{equation}

\noindent
while $\widetilde{\phi}_{ij}$ are determined through $\widetilde{\chi}_{ij}$ as follows:
\begin{equation}
\widetilde{\phi}_{11}=-{1 \over 2m_f}[(i+1)\epsilon \partial_\xi\widetilde{\chi}_{12}-i k_z \partial_\tau\widetilde{\chi}_{11}],
\label{eq:eq22}
\end{equation}

\begin{equation}
\widetilde{\phi}_{12}=-{1 \over 2m_f}[(i-1)\epsilon \partial_\xi\widetilde{\chi}_{11}+i k_z \partial_\tau\widetilde{\chi}_{12}-gA\widetilde{\chi}_{21}],
\label{eq:eq23}
\end{equation}

\begin{equation}
\widetilde{\phi}_{21}=-{1 \over 2m_f}[(i+1)\epsilon \partial_\xi\widetilde{\chi}_{22}-i k_z \partial_\tau\widetilde{\chi}_{21}-gA\widetilde{\chi}_{12}],
\label{eq:eq24}
\end{equation}

\begin{equation}
\widetilde{\phi}_{22}=-{1 \over 2m_f}[(i-1)\epsilon \partial_\xi\widetilde{\chi}_{21}+i k_z \partial_\tau\widetilde{\chi}_{22}].
\label{eq:eq25}
\end{equation}

\noindent
Equations~(\ref{eq:eq18}-\ref{eq:eq21}) can be consistently reduced, using $\widetilde{\chi}_{11}=\widetilde{\chi}_{22}$ and $\widetilde{\chi}_{21}= i \widetilde{\chi}_{12}$, to the following two equations:
\begin{equation}
i \partial_t \widetilde{\chi}_{11} + {\nabla^2 \over 2m_f}\widetilde{\chi}_{11}={1-i \over 2m_f}g\epsilon^2f(\tau)\partial_\xi(\overline{A}(\xi)\widetilde{\chi}_{12}),
\label{eq:eq26}
\end{equation}

\begin{eqnarray}
i \partial_t \widetilde{\chi}_{12}+{\nabla^2 \over 2m_f}\widetilde{\chi}_{12}=&{ k_z \over 2m_f}g\epsilon(\partial_\tau f(\tau))\overline{A}(\xi)\widetilde{\chi}_{12} -\epsilon^2{1+i \over 2m_f}g\overline{A}(\xi)f(\tau) \partial_\xi\widetilde{\chi}_{11} \nonumber\\
&+{g^2 \epsilon^2\over 2m_f}\overline{A}^2(\xi)f^2(\tau)\widetilde{\chi}_{12},
\label{eq:eq27}
\end{eqnarray}

\noindent
where $f(\tau)=\cos(\tau)$ and $\overline{A}(\xi)={2 \alpha \over \sqrt{3}g}\tanh(\alpha \xi)$.
Without loss of generality we can choose $k_z=0$ (using the rest frame of the massive gauge field as reference frame) to further simplify the above expressions. Furthermore, in order to allow for non-trivial dynamics in the fermionic field,
the corresponding mass $m_f$ has to be small (of order $\mathcal{O}(\epsilon^2)$) as compared to the gauge field mass. In this case, writing $m_f= \epsilon^2 m_0$, we obtain the following system of two equations:
\begin{equation}
i \partial_t \widetilde{\chi}_{11} + {1 \over m_0}\partial^2_\xi\widetilde{\chi}_{11}={1-i \over 2m_0}gf(\tau)\partial_\xi(\overline{A}(\xi)\widetilde{\chi}_{12}),
\label{eq:eq28}
\end{equation}

\begin{equation}
i \partial_t \widetilde{\chi}_{12}+{1 \over m_0}\partial^2_\xi\widetilde{\chi}_{12}=
-{1+i \over 2m_0}g\overline{A}(\xi)f(\tau) \partial_\xi\widetilde{\chi}_{11}
+{g^2 \over 2m_0}\overline{A}^2(\xi)f^2(\tau)\widetilde{\chi}_{12},
\label{eq:eq29}
\end{equation}
where $m_0$ is a mass scale of the order of $m_g$.

\noindent
Let us now introduce the length scale $\xi_0$ and the time scale $\sigma=m_0 \xi_0^2$ to express
Eqs.~(\ref{eq:eq28}-\ref{eq:eq29}) in a dimensionless form. In these units, the dimensionless frequency of the oscillating YM-field becomes: $\omega_0=m_g m_0 \xi_0^2$. It also straightforward to define dimensionless variables $\xi=\xi_0 \rho$ and $t=\sigma \widetilde{\tau}$. In these variables, we seek for solutions of the system (\ref{eq:eq28}-\ref{eq:eq29}) having the form:
\begin{equation}
\widetilde{\chi}_{11}(\rho,\widetilde{\tau})=e^{-i \lambda m_0 \xi_0^2 \widetilde{\tau}} F(\rho,\widetilde{\tau})~~~;~~~\widetilde{\chi}_{12}=e^{-i \lambda m_0 \xi_0^2 \widetilde{\tau}} G(\rho,\widetilde{\tau}),
\label{eq:eq30}
\end{equation}
where $F$ and $G$ are slowly-varying functions of $\widetilde{\tau}$, while $\lambda$ is the energy eigenvalue. In this limit, Eqs.~(\ref{eq:eq28}-\ref{eq:eq29}) become:
\begin{equation}
\lambda m_0 \xi^2_0 F(\rho,\widetilde{\tau})+ \partial^2_{\rho} F(\rho,\widetilde{\tau})={1-i \over \sqrt{3}}\alpha \xi_0 \cos(\omega_0 \widetilde{\tau})\partial_{\rho}[G(\rho,\widetilde{\tau})\tanh(\alpha \xi_0 \rho)],
\label{eq:eq31}
\end{equation}

\begin{eqnarray}
\lambda m_0 \xi^2_0 G(\rho,\widetilde{\tau})+ \partial^2_{\rho} G(\rho,\widetilde{\tau})&=-{1+i \over \sqrt{3}}\alpha \xi_0 \cos(\omega_0 \widetilde{\tau})\tanh(\alpha \xi_0 \rho)\partial_{\rho} F(\rho,\widetilde{\tau})\nonumber\\
&+{2\over 3}(\alpha \xi_0)^2\cos^2(\omega_0 \widetilde{\tau})\tanh^2(\alpha \xi_0 \rho)G(\rho,\widetilde{\tau}).
\label{eq:eq32}
\end{eqnarray}

\noindent
For $\omega_0=m_g m_0 \xi_0^2 \gg 1$, Eqs.~(\ref{eq:eq31}-\ref{eq:eq32}) can be integrated with respect to $\widetilde{\tau}$ over a period $T=\frac{2 \pi}{m_g m_0 \xi_0^2}$ since in this time interval $F$ and $G$ are practically constant. Following this procedure, Eqs.~(\ref{eq:eq31}-\ref{eq:eq32}) decouple and obtain the following form:
\begin{equation}
\lambda m_0 \xi^2_0 F(\rho,\widetilde{\tau})+ \partial^2_\rho F(\rho,\widetilde{\tau})=0,
\label{eq:eq33}
\end{equation}

\begin{equation}
(\lambda m_0 \xi^2_0-{1\over 3 } (\alpha \xi_0)^2)G(\rho,\widetilde{\tau})+ \partial^2_\rho G(\rho,\widetilde{\tau})+{1\over 3 } (\alpha \xi_0)^2{1 \over \cosh^2(\alpha \xi_0 \rho )}G(\rho,\widetilde{\tau})=0,
\label{eq:eq34}
\end{equation}

\noindent
allowing as a solution a fermionic state which is bound in the $\xi$ direction and has the form \cite{Landau}:
\begin{eqnarray}
F(\rho,\widetilde{\tau})&=& F(\xi)=0, \nonumber\\
G(\rho,\widetilde{\tau})=G(\xi)=N\Big({1 \over \cosh^2(\alpha \xi)}\Big)^{s}&,&s=\frac{1}{4}\left(\sqrt{1+\frac{4}{3}}-1\right)\approx 0.1319, \nonumber\\
\label{eq:eq35}
\end{eqnarray}

\noindent
where $N$ is a normalization constant.
The state (\ref{eq:eq35}) resembles the Landau levels of a particle in an external magnetic field in quantum electrodynamics. In the YM case under consideration, the magnetic field is generated by the term proportional to $\overline{A}^2(\xi)$ in Eq.~(\ref{eq:eq29}). The difference here is that we have a single level independently of the strength of the external Yang-Mills field. In addition, the Dirac particle is trapped only in the $\xi$-direction, where the external field is also localized. It should be noticed that the condition $m_g m_0 \xi_0^2 \gg 1$, necessary for the existence of the solution (\ref{eq:eq35}), can be justified by either using a large $\xi_0$ value or a large $m_g$ value (or both).

It is illuminating to give an example of the energy and length scales involved in this solution. Assuming a gauge field mass of $500$~MeV and a much smaller fermionic mass i.e., of order of $\mathcal{O}(5~{\rm MeV})$, we find that the $SU(2)$ charged fermions are trapped in a region of radius of $\approx 150~fm$ in the $(x,y)$-plane with energy eigenvalue $\approx 1.5~MeV$ for an external field of amplitude $6~MeV$. It must be noted that for this choice of parameter values the non-relativistic approximation is valid within an error of 15\% estimated by the relative magnitude of the first relativistic correction term. In Fig.~3 we show the effective potential responsible for the trapping of the Dirac particle using the
above mentioned parameter values. The dashed line indicates the energy $\lambda=s F_0^2$ of the associated bound state in the $\xi$-space. The fact that this state is very close to the continuum threshold explains the absence of a second bound state. In Fig.~4 we show the $\xi$-dependent wave function corresponding to the bound state displayed in Fig.~3. The broad spatial extension of this state is attributed to the small exponent in Eq.~(\ref{eq:eq35}).

\begin{figure}[ht]
\includegraphics[scale=0.65]{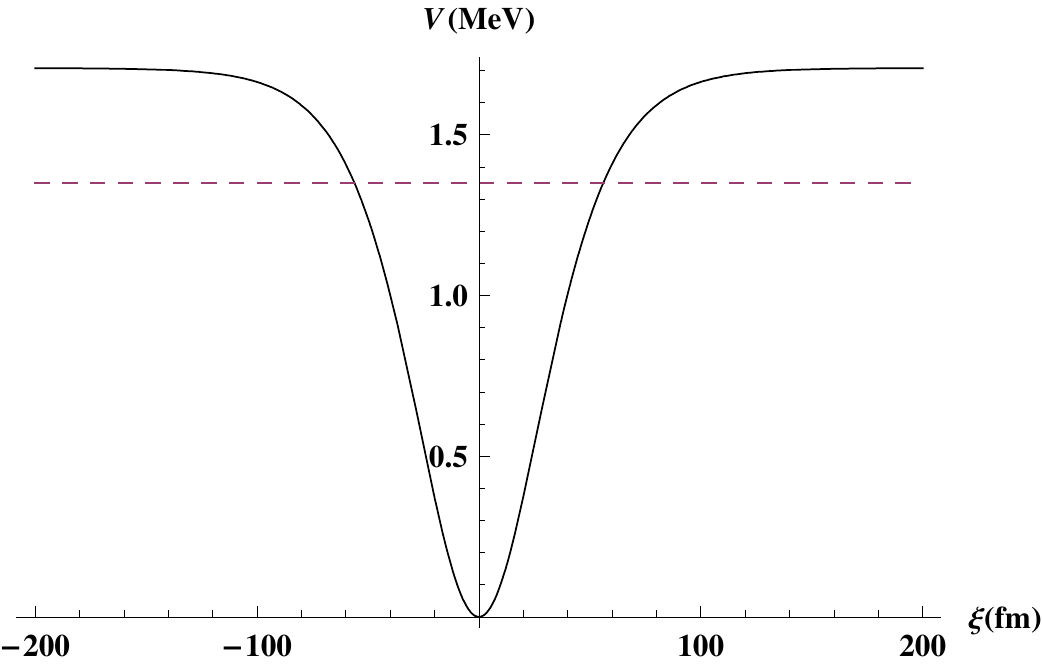}
\caption{The effective potential responsible for the trapping of a Dirac particle with $SU(2)$ charge emerging from the a time-dependent external Yang-Mills field of the type shown in Fig.~1. The parameters values used are: $k_0=500~MeV$, $F_0=3.2~MeV^{1/2}$. The dashed line indicates the energy of the bound state.
\label{fig3}}
\end{figure}

\begin{figure}[ht]
\includegraphics[scale=0.65]{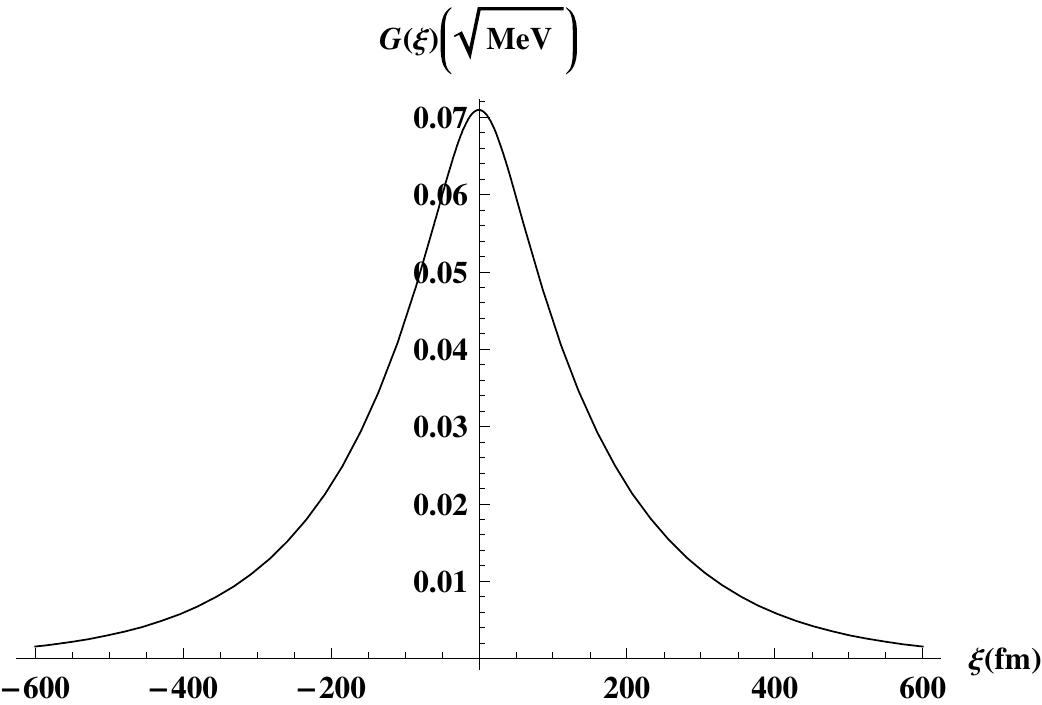}
\caption{The $\xi$-dependent normalized wavefunction of the bound state shown in Fig.~3 calculated using the same parameter values.
\label{fig4}}
\end{figure}

\section{Conclusions and discussion}

We have investigated classical solutions of the $SU(2)$ massive Yang-Mills equations in the framework of multiple scale perturbation theory. Due to the presence of the mass term, conformal symmetry is explicitly broken and the Coleman theorem does not apply \cite{Coleman77}. Therefore, the YM dynamics in this case admit soliton-like solutions localized in a subspace of the transverse space.

Such solutions of the Yang-Mills field break both Lorentz and gauge invariance in higher orders of the perturbation expansion, in consistency with the presence of a mass term as well as the appearance of partial localization. Dirac fermions with non-vanishing $SU(2)$ charge, when exposed in an external YM field having the form of these soliton-like solutions, become trapped in a similar way as electrons in a transverse magnetic field (Landau levels). However, the trapping of the $SU(2)$ colored fermions is a pure dynamical effect occurring in the non-adiabatic limit of very fast oscillations of the external YM field, and occurs only along the ($x+y$)-direction.

Our analysis reveals a mechanism for the occurrence of localized fermionic states with $SU(2)$ charge based on the interaction with a massive Yang-Mills field. The simplifying assumptions made in our approach (two non-vanishing equal components of the gauge field at the leading order) may restrict the profile of the found solutions allowing, on the other hand, for an analytical treatment. Despite this restriction, the main ingredients of the present study could be used as a guide to obtain more general inhomogeneous classical solutions of the massive $SU(2)$ field. However, such a task is a subject for future investigations.

\begin{acknowledgements}
We thank N. G. Antoniou, E. G. Floratos and A. Tsapalis for helpful discussions.
This work was partially supported by the Special Account for Research Grants of the
University of Athens.
\end{acknowledgements}

\appendix
\section{}
Using the classification of the gauge fields in orders of $ \epsilon^3$ cf.~Eqs.~(\ref{eq:eq5}), we can write the equations of motion for the components $A_1^1(1)$, $A_2^2(1)$ as follows:
\begin{eqnarray}
\mathcal{O}(\epsilon):& (\Box_0 + m_g^2 + \partial_{k_0}^2)A_k^k(1)=0 \label{eq:eqA1a},\\
\mathcal{O}(\epsilon^2):&(\Box_0 +m_g^2 +\partial_{k_0}^2) A_k^k(2) +  2(\partial_{\mu_0} \partial^{\mu_1}+\partial_{k_0} \partial_{k_1})A_k^k(1)=0 \label{eq:eqA1b},\\
\mathcal{O}(\epsilon^3):&(\Box_0 + m_g^2 + \partial_{k_0}^2 )A_k^k(3) + 2(\partial_{\mu_0} \partial^{\mu_1} +  \partial_{k_0} \partial_{k_1})A_k^k(2)+i k_0 \partial_{k_1}A^k_0(2) \nonumber\\
& +i k_z \partial_{k_1}A^k_3(2)+(\Box_1 + \partial_{k_1}^2+ 2\partial_{\mu_0} \partial^{\mu_2} + 2 \partial_{k_0} \partial_{k_2} )A_k^k(1) +g^2 S_k=0,\phantom{aaaaaa}\label{eq:eqA1c}
\end{eqnarray}
where $S_1\equiv A_1^1(1)A_2^2(1)A_2^2(1)$ and $S_2 \equiv A_2^2(1)A_1^1(1)A_1^1(1)$.

The non-diagonal equations, as well as the equations for the case $\nu=3,a=3$, are obtained in a similar way and their consistency with the choice in Eq.~(\ref{eq:eq5}) implies the following condition:
\begin{equation}
i k_0 \partial_{k_1} A^k_0(2) +i \partial_{k_1}A^k_3(2)=-\partial^2_\xi A^k_k(1), \label{eq:eqA2}
\end{equation}
for every $k=1,2$. Thus, Eq.~({\ref{eq:eqA1c}) becomes:
\begin{eqnarray}
(\Box_0 + m_g^2 + \partial_{k_0}^2 )A_k^k(3) + 2(\partial_{\mu_0} \partial^{\mu_1} +  \partial_{k_0} \partial_{k_1})A_k^k(2)-\partial^2_{\xi} A^k_k(1) \nonumber\\
+(\Box_1 + \partial_{k_1}^2+2\partial_{\mu_0} \partial^{\mu_2} + 2\partial_{k_0} \partial_{k_2} )A_k^k(1) +g^2 S_k=0 \quad.
\label{eq:eqA3}
\end{eqnarray}
In Eq.~(\ref{eq:eqA3}) the fields $A_1^1$ and $A_2^2$ are still coupled due to the presence of the nonlinear term $S_k$; nevertheless, we can readily resolve this problem by assuming that $ A_1^1 \equiv A_2^2$.
Equation~(\ref{eq:eqA1a}) reveals the dependence on the normal scales $x_\mu$ (in the first order of the perturbation expansion) of the gauge field, as it admits a harmonic solution for $A_k^k(1)$ of the form:
\begin{equation}
A_k^k(1)=f_1^1(X_{\mu_1},X_{\mu_2},...)e^{-i \tau}+c.c. \quad, \tau=k_0 t-k_z z,k_0^2=k_z^2+m_g^2.
\label{eq:eqA4}
\end{equation}
The function $f_1^1(X_{\mu_1},X_{\mu_2},...)\equiv f(1)$, which is for the moment an arbitrary complex function will be consistently determined by solving the equations arising at higher orders of $\epsilon$.

Next, considering Eq.~(\ref{eq:eqA1b}), it is clear that the homogenous part of the solution is similar to the one in
Eq.~(\ref{eq:eqA4}), due to the fact that the linear operator in (\ref{eq:eqA1b}) and in (\ref{eq:eqA1a}) are identical. As a result, the term $2(\partial_{\mu_0} \partial^{\mu_1}+\partial_{k_0} \partial_{k_1})A_k^k(1)$ is secular, as
$A_k^k(2)$ will contain terms of the form $\tau e^{-i \tau}$.

The condition for nonsecularity, namely $2(\partial_{\mu_0} \partial^{\mu_1}+\partial_{k_0} \partial_{k_1})A_k^k(1)=0$, leads to the following two equations [valid at order $\mathcal{O}(\epsilon^2)]$:
\begin{eqnarray}
(\Box_0 +m_g^2 +\partial_{k_0}^2) A_k^k(2)=0,& \label{eq:eqA5a} \\
(\partial_{\mu_0} \partial^{\mu_1}+\partial_{k_0} \partial_{k_1})A_k^k(1)=0.&
\label{eq:eqA5b}
\end{eqnarray}
Since $A_k^k(1)$ does not depend on $x$ and $y$ [cf.~Eq.~(\ref{eq:eqA4})], one has $\partial_{k_0} \partial_{k_1}A_k^k(1)=0$ for $k=1,2$; furthermore, the condition $\partial_{\mu_0} \partial^{\mu_1}A_k^k(1)=0$,
introduces an important restriction for the function $f(1)$ in Eq.~(\ref{eq:eqA4}): it is necessary to assume that $f(1)=f_k^k(X_1,\Psi_1;X_{\mu_2}..)$, i.e., $f(1)$ is independent of $X_{0_1}\equiv T_1$ and $X_{3_1} \equiv Z_1$, a fact which sustains the decomposition of space-time in two inequivalent subspaces, as mentioned in Sec.~II.

Finally, Eq.~(\ref{eq:eqA3}) decomposes in three independent equations. The first of them reads:
\begin{equation}
(\Box_0 + m_g^2 + \partial_{k_0}^2 )A_k^k(3)+ g^2~({\rm nonsecular~part~of~S_k})=0.
\label{eq:eqA6}
\end{equation}
The remaining two equations are found by eliminating all secular terms producing divergence of $A_k^k(3)$ in Eq.~(\ref{eq:eqA3}). This way, we have:
\begin{equation}
(\partial_{\mu_0} \partial^{\mu_1} +  \partial_{k_0} \partial_{k_1})A_k^k(2)=0, \label{eq:eqA6b}
\end{equation}
which is treated in the same way as Eq.~(\ref{eq:eqA5b}) for the $A_k^k(1)$ field, and
\begin{eqnarray}
(\Box_1 + 2\partial_{\mu_0} \partial^{\mu_2}+\partial^2_{k_1})f^k_k(1)e^{-i \tau} -\partial^2_{k_1}f^k_k(1)e^{-i \tau}+ \nonumber \\
+g^2~({\rm secular~part~of~S_k})=0.
\label{eq:eqA6c}
\end{eqnarray}

Our assumption that $f^1_1(1) \equiv f^2_2(1)$ implies that $S_k = (A_1^1(1))^3=(A_2^2(1))^3$ and, as a result, Eq.~(\ref{eq:eqA6c}) should be of the same form for $k=1,2$. This requirement is satisfied if $\partial_{X_1} \equiv \partial_{\Psi_1}$ and $f(1) \equiv f_k^k(X_1+\Psi_1; X_{\mu_2},..)$. Consequently, Eq.~(\ref{eq:eqA6c}) is reduced to the form:
\begin{equation}
-2\partial_{\xi}^2 f(1) - 2i k_0  \partial_{T_2}f_k^k(1)   + 3g^2 f_k^k(1)|f_k^k(1)|^2 =0, \label{eq:eqA7}
\end{equation}
where $A_k^k(1)= f_k^k(1)\exp(-i\tau)~+~c.c$ .

As far as Eq.~(\ref{eq:eqA6c}) is concerned, it is important to note that the second term is the contribution of the non-diagonal terms [cf.~Eqs.~(\ref{eq:eqA1c}) and (\ref{eq:eqA2})]. Note that Eq.~(\ref{eq:eqA7}) is actually the NLS equation presented in Sec. II (see Eq.~(\ref{eq:eq12})).

\section{}
We start by rewriting Eqs.~(\ref{eq:eq15}-\ref{eq:eq16}) in the following form:
\begin{eqnarray}
i \gamma^0 \partial_0 \Psi_1+ i {\gamma} \cdot \nabla \Psi_1 -m_f \Psi_1={1 \over 2}g \epsilon \overline{A}(\xi)\cdot \cos\tau (\gamma^1- i \gamma^2)\Psi_2,
\label{eq:eqB1a}
\end{eqnarray}
\begin{eqnarray}
i \gamma^0 \partial_0 \Psi_2+ i {\gamma} \cdot \nabla \Psi_2 -m_f \Psi_2={1 \over 2}g \epsilon \overline{A}(\xi)\cdot \cos\tau (\gamma^1+ i \gamma^2)\Psi_1,
\label{eq:eqB1b}
\end{eqnarray}
where $\Psi_1$ and $\Psi_2$ are the two components of the bispinor $\Psi$ defined in Eq.~(\ref{eq:eq17}). In the following, we will apply the standard procedure \cite{Bjorken78} in order to obtain the non-relativistic limit of Eqs.~(\ref{eq:eqB1a}-\ref{eq:eqB1b}). The necessity of this emerges by the violation of the covariance of the gauge field which we have imposed. Eventually, it is consistent to study the non-relativistic case.

Taking into account that $ \left( \begin{array}{c}\chi_i \\ \phi_i \end{array} \right)=e^{-i m_f t} \left( \begin{array}{c} \widetilde{\chi_i} \\ \widetilde{\phi}_i \end{array}\right)$, for $i=1,2$, Eq.~(\ref{eq:eqB1a}) transforms into
\begin{equation}
i \partial_t \left( \begin{array}{c} \widetilde{\chi}_1 \\ -\widetilde{\phi}_1 \end{array} \right) -2m_f \left( \begin{array}{c}0 \\ \widetilde{\phi}_1 \end{array} \right)+i {\gamma} \cdot \nabla\left( \begin{array}{c}\widetilde{\chi}_1 \\ \widetilde{\phi}_1 \end{array} \right)=\epsilon g \overline{A}(\xi)\cdot \cos\tau  \left( \begin{array}{c} 0 \\ \widetilde{\phi}_{21} \\ 0 \\ -\widetilde{\chi}_{21} \end{array} \right).
\label{eq:eqB2}
\end{equation}
From Eq.~(\ref{eq:eqB2}) we obtain the following equations for the doublets $\widetilde{\chi}_1 $ and $ \widetilde{\phi}_1$ of the field $\Psi_1$:

\begin{equation}
i \partial_t \widetilde{\chi}_1+i {\sigma}\cdot \nabla \widetilde{\phi}_1=\epsilon g \overline{A}(\xi)\cdot \cos\tau  \left( \begin{array}{c} 0 \\ \widetilde{\phi}_{21} \end{array}\right), \label{eq:eqB3a}
\end{equation}
\begin{equation}
-i \partial_t \widetilde{\phi}_1-2m_f \widetilde{\phi}_1- i {\sigma}\cdot \nabla \widetilde{\chi}_1=-\epsilon g \overline{A}(\xi)\cdot \cos\tau  \left( \begin{array}{c} 0 \\ \widetilde{\chi}_{21} \end{array}\right), \label{eq:eqB3b}
\end{equation}
and similarly for the field $\Psi_2$:
\begin{equation}
i \partial_t \widetilde{\chi}_2 + i {\sigma}\cdot \nabla \widetilde{\phi}_2=\epsilon g \overline{A}(\xi)\cdot \cos\tau  \left( \begin{array}{c} \widetilde{\phi}_{12}  \\ 0 \end{array}\right), \label{eq:eqB4a}
\end{equation}
\begin{equation}
-i \partial_t \widetilde{\phi}_2-2m_f \widetilde{\phi}_2- i {\sigma}\cdot \nabla \widetilde{\chi}_2=-\epsilon g \overline{A}(\xi)\cdot \cos\tau  \left( \begin{array}{c} \widetilde{\chi}_{12}  \\ 0 \end{array}\right), \label{eq:eqB4b}
\end{equation}
where $\widetilde{\phi}_1 $ and $\widetilde{\phi}_2$ are slowly varying functions of time, while $\widetilde{\chi}_{ij} \sim e^{-i \lambda m_0 \xi_0^2 \tau} F(y,\tau)$ or $G(y,\tau)$, with $F$, $G$ being slowly varying functions of time as well, and $\omega_0 \gg 1$. Using the relations
\begin{equation}
\widetilde{\phi}_1=-{1 \over 2m_f }\left[ i {\sigma}\cdot \nabla \widetilde{\chi}_1 - \epsilon g \overline{A}(\xi)\cdot \cos\tau \left( \begin{array}{c} 0 \\ \widetilde{\chi}_{21}\end{array}\right)\right]  \label{eq:eqB5a}
\end{equation}
\begin{equation}
\widetilde{\phi}_2=-{1 \over 2m_f }\left[ i {\sigma}\cdot \nabla \widetilde{\chi}_2 -\epsilon g \overline{A}(\xi)\cdot \cos\tau \left( \begin{array}{c} \widetilde{\chi}_{12} \\ 0 \end{array}\right)\right] \label{eq:eqB5b}
\end{equation}
 equation (\ref{eq:eqB3a}) becomes
 $$i \partial_t \widetilde{\chi}_1+i {\sigma}\cdot \nabla(-{1 \over 2m_f })\left[ i {\sigma}\cdot \nabla \widetilde{\chi}_1 - \epsilon g \overline{A}(\xi)\cdot \cos\tau \left( \begin{array}{c} 0 \\ \widetilde{\chi}_{21}\end{array}\right)\right] =\epsilon g \overline{A}(\xi)\cdot \cos\tau  \left( \begin{array}{c} 0 \\ \widetilde{\phi}_{21} \end{array}\right),
 $$
 and since $ ({\sigma}\cdot \nabla)^2=\nabla^2$ we have
\begin{equation}
i \partial_t \widetilde{\chi}_1+ {\nabla^2 \over 2m_f}\widetilde{\chi}_1  + i {1 \over 2m_f }   \epsilon g {\sigma}\cdot \nabla \left(\overline{A}(\xi)\cdot \cos\tau \left( \begin{array}{c} 0 \\ \widetilde{\chi}_{21}\end{array}\right)\right)=\epsilon g \overline{A}(\xi)\cdot \cos\tau  \left( \begin{array}{c} 0 \\ \widetilde{\phi}_{21} \end{array}\right)
\label{eq:eqB6}
\end{equation}
while for the $\widetilde{\chi}_2$ component, similarly we have
\begin{equation}
i \partial_t \widetilde{\chi}_2+ {\nabla^2 \over 2m_f}\widetilde{\chi}_2  +  i {1 \over 2m_f }   \epsilon g {\sigma}\cdot \nabla \left(\overline{A}(\xi)\cdot \cos\tau \left( \begin{array}{c} \widetilde{\chi}_{12}  \\ 0 \end{array}\right)\right)=\epsilon g \overline{A}(\xi)\cdot \cos\tau  \left( \begin{array}{c} \widetilde{\phi}_{12}   \\ 0 \end{array}\right) \label{eq:eqB7}
\end{equation}
where $\overline{A}(\xi)={2 \alpha \over \sqrt{3}g}\tanh(\alpha \xi)$, $\xi=\epsilon (x+y)$ and $\alpha=\sqrt{{k_0 \over 2}}F_0$. Finally, we expand Eq.~(\ref{eq:eqB6}) and (\ref{eq:eqB7}) in their components resulting in
Eqs.~(\ref{eq:eq18}-\ref{eq:eq21}) for the $\widetilde{\chi}_{ij}$ fields.\\

\end{document}